\documentclass[aps,prl, superscriptaddress, showpacs,twocolumn]{revtex4-1}
\pdfoutput=1

\usepackage{graphicx, color}
\usepackage{dcolumn}
\usepackage{bm}
\usepackage{amsmath}
\usepackage{amssymb}
\usepackage{verbatim}
\usepackage{appendix}
\usepackage{natbib}
\usepackage{float}
\usepackage{gensymb}

\begin{document}


\title{Room-Temperature \textit{in situ} Nuclear Spin Hyperpolarization from Optically-Pumped Nitrogen Vacancy Centers in Diamond}


\author{Jonathan P. King}
\affiliation{Department of Chemistry, University of California, Berkeley, California 94720}
\affiliation{Materials Sciences Division, Lawrence Berkeley National Laboratory, Berkeley, California 94720}
\author{Keunhong Jeong}
\affiliation{Department of Chemistry, University of California, Berkeley, California 94720}
\affiliation{Materials Sciences Division, Lawrence Berkeley National Laboratory, Berkeley, California 94720}
\author{Christophoros C. Vassiliou}
\affiliation{Department of Chemistry, University of California, Berkeley, California 94720}
\affiliation{Materials Sciences Division, Lawrence Berkeley National Laboratory, Berkeley, California 94720}
\author{Chang S. Shin}
\affiliation{Department of Chemistry, University of California, Berkeley, California 94720}
\affiliation{Materials Sciences Division, Lawrence Berkeley National Laboratory, Berkeley, California 94720}
\author{Ralph H. Page}
\affiliation{Department of Chemistry, University of California, Berkeley, California 94720}
\author{Claudia E. Avalos}
\affiliation{Department of Chemistry, University of California, Berkeley, California 94720}
\affiliation{Materials Sciences Division, Lawrence Berkeley National Laboratory, Berkeley, California 94720}
\author{Hai-Jing Wang}
\affiliation{Department of Chemistry, University of California, Berkeley, California 94720}
\affiliation{Materials Sciences Division, Lawrence Berkeley National Laboratory, Berkeley, California 94720}
\author{Alexander Pines}
\email[]{pines@berkeley.edu}
\affiliation{Department of Chemistry, University of California, Berkeley, California 94720}
\affiliation{Materials Sciences Division, Lawrence Berkeley National Laboratory, Berkeley, California 94720}



\date{\today}

\begin{abstract}
We report bulk, room-temperature hyperpolarization of $^{13}$C nuclear spins observed via high-field nuclear magnetic resonance (NMR). The hyperpolarization is achieved by optical  pumping (OP) of nitrogen vacancy defect centers in diamond accompanied by dynamic nuclear polarization (DNP). The technique harnesses the large optically-induced spin polarization of NV$^-$ centers at room temperature, which is many orders of magnitude greater than thermal equilibrium polarization and typically achievable only at sub-Kelvin temperatures. Transfer of the spin polarization to the $^{13}$C nuclear spins is accomplished via a combination of OP and microwave irradiation. The OP/DNP is performed at 420 mT, where inductive detection of NMR is feasible, in contrast to the typically exploited level anticrossing regimes at 100 mT and 50 mT. Here, we report a bulk nuclear spin polarization of 6$\%$. This polarization was generated \textit{in situ} and detected with a standard, inductive NMR probe without the need for sample shuttling or precise crystal orientation. Hyperpolarization via OP/DNP should operate at arbitrary magnetic fields, enabling orders of magnitude sensitivity enhancement for NMR of solids and liquids at ambient conditions. 

\end{abstract}


\maketitle
Nuclear magnetic resonance spectroscopy (NMR) and imaging (MRI) are indispensable techniques in fields reaching from chemistry and materials to biology and medicine. The appealing non-destructive nature and broad range of applications for these techniques notwithstanding, they are subject to several  limitations. Significant among these limitations is the low NMR signal sensitivity of the weakly magnetized nuclear spins at room temperature, where the polarization can be less than one part per million. An intensive, ongoing goal of contemporary magnetic resonance is the development of methods to produce non-thermal states of nuclear spin hyperpolarization. Examples include: spin exchange optical pumping of noble gasses with alkali vapors \cite{raftery,happer,schroeder,carver}; optical pumping of semiconductors \cite{jefftrends, hayesreview}; parahydrogen induced polarization \cite{mewis,weitekamp,bargon}; low temperature dynamic nuclear polarization (DNP) \cite{griffin,prisner,larsen,bodenhausen}; chemically induced DNP \cite{goez}; and optical pumping with DNP of excited triplet states in organic solids \cite{triplet}. 

Nitrogen vacancy (NV$^-$) centers in diamond, with their optically-polarized spin states and optical spin readout, have provided a means to detect nuclear spins with high sensitivity and spatial resolution resulting, most recently, in the detection of a single proton spin \cite{rugar,devince,degen,jelezko,sushkov}. Nuclear spins hosted within the diamond lattice have been hyperpolarized using level anti-crossings that occur at specific crystal orientations and magnetic field strengths \cite{fischerensemble,haijing}. Evidence of nuclear spin hyperpolarization of proximate $^{13}$C spins was deduced from optically-detected magnetic resonance (ODMR) spectra at the level anticrossing fields and subsequently confirmed at a value of approximately $0.5\%$ in bulk by  shuttling the diamond sample to a higher magnetic field for NMR detection \cite{frydmanbulk}. These techniques were then extended to low fields away from the anticrossing and to arbitrarily oriented NV$^-$ centers using microwave irradiation \cite{alvarez}. Bulk $^{13}$C polarization has been generated at high field and low temperature and attributed to the coupling of the nuclear spins to the dipolar energy reservoir of the NV$^-$ ensemble \cite{king};  the precise mechanism remains unclear. Furthermore, a recent proposal has suggested the use of shallow NV$^-$ centers as a source for direct polarization transfer to a surrounding liquid \cite{carlos}. Particularly desirable would be a general method to produce hyperpolarization \textit{in situ} under the same magnetic field and temperature conditions as the NMR experiment using an inert, non-toxic, and easily separated source. In our method, optical pumping of diamonds coupled with DNP, under ambient conditions, obviates the need for cryogenic temperatures, sample shuttling, and precise crystal orientation and magnetic field strengths, thereby providing such a general method for high-sensitivity NMR at arbitrary field strength.

The electronic ground state of the NV$^-$ center is a spin-1 triplet with a zero-field splitting $D=2.87$ GHz between the $m_s=\pm1$  and $m_s=0$ states. The $m_s=0$ state is preferentially populated via optical pumping with a 532 nm laser and spin-dependent nonradiative decay rates. Applying a magnetic field along the defect symmetry axis lifts the degeneracy of the $m_s=\pm1$ states and gives rise to two distinct magnetic resonance transitions observable by ODMR. ODMR relies on a reduction in the fluorescence intensity induced by depopulation of the $m_s=0$ state (Fig. 1). In our experiment, $^{13}$C spins in a 4.5 mg diamond are hyperpolarized by DNP using an optically-polarized microwave transition. A strong NMR signal was observed from this natural isotopic abundance sample after accumulating 60 scans with a repetition time of 60 seconds (Fig. 2). For comparison and calibration, after accumulating 12676 scans with a repetition time of 10 ms, a 10 $\mu$l sample of Gd(III)-doped liquid dimethylsulfoxide (DMSO) enriched to 99\% $^{13}$C gave an NMR signal lower than the diamond by a factor of $\sim$12. From the ratio of the numbers of $^{13}$C nuclei in the diamond  and DMSO samples (0.015), the number of scans needed for each, and the ratio of signal amplitudes, the maximum bulk $^{13}$C polarization in the diamond is estimated to be 6\%. The nuclear spin polarization builds up over several minutes (Fig. 3b). We attribute the dynamics to the coupled processes of DNP of nuclear spins proximate to the NV$^-$ centers, nuclear spin diffusion to the bulk material \cite{reynhardt,abragam}, and spin-lattice relaxation. This process is shown schematically in Fig. 3a. The mechanisms for DNP using paramagnetic impurities in diamond are well known \cite{reynhardt} and summarized in the supplementary material. The OP/DNP process is expected to be effective at arbitrary orientations of the NV$^-$ defects, since the process depends on a matching of the microwave frequency to a given transition, rather than a precise field strength and orientation. To test this idea, the sample was rotated 90$\degree$ around the axis perpendicular to the laser and magnetic field. This ensured that no NV$^-$ centers were aligned with the magnetic field. An ODMR signal was found at 14,402 MHz, which corresponds to an NV$^-$ misalignment of 14$\degree$ from the field, and DNP data were collected in this region (Fig. 3c), showing $^{13}$C spin polarization approaching 2\%. The effectiveness of the DNP for misaligned NV$^-$ centers will be critical for the extension of this technique to randomly oriented powders.

 
 
 \begin{figure}
 \includegraphics[width=0.5\textwidth]{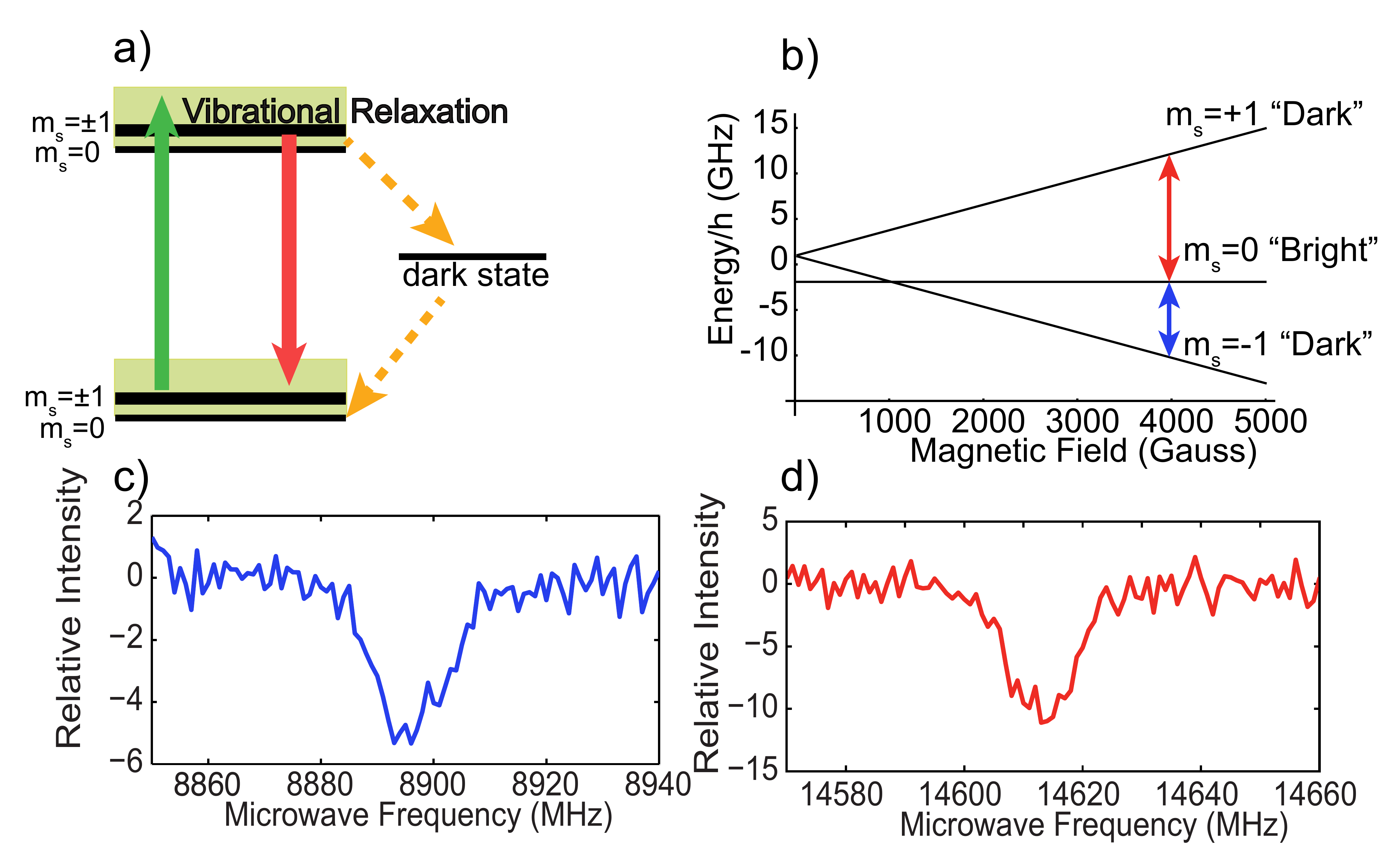}%
 \caption{\label{levels}a) Energy levels and transitions for an NV$^-$ center in diamond. Optical pumping with green light at 532 nm induces transitions from the ground state spin-1 triplet to the excited triplet state.  Subsequent to vibrational relaxation, fluorescence  is detected in the red and near-IR. Spin conserving optical transitions and spin-dependent non-radiatve intersystem crossings lead to a preferential population of the $m_s=0$ ground state producing electron spin hyperpolarization of the NV$^-$ center. b) Application of a magnetic field aligned along the NV$^-$ axis lifts the degeneracy of the $m_s=\pm1$ states, yielding two transitions that can be driven with microwave irradiation. The two transitions,  c) between $m_s=0$ and $m_s=-1$ and d) between $m_s=0$ and $m_s=+1$, are observed by optically-detected magnetic resonance (ODMR) through a reduction in the fluorescence intensity caused by a depletion of the ground $m_s=0$ state.}
 \end{figure}
 
   \begin{figure}[H]
 \includegraphics[width=0.5\textwidth]{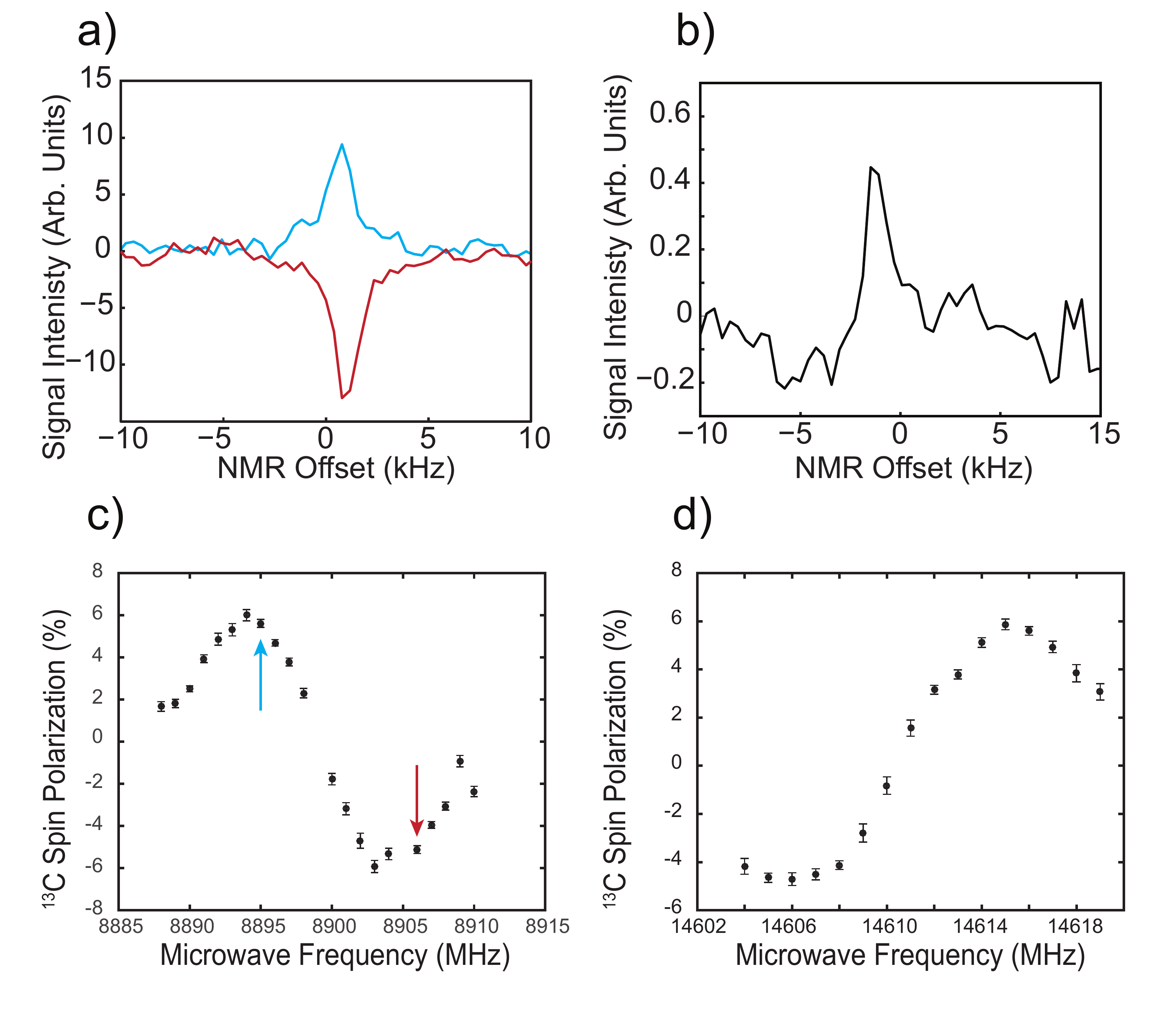}%
 \caption{\label{spectra}a) NMR spectra of $^{13}$C at natural abundance in diamond after the accumulation of 60 scans under DNP for 60 seconds at 8895 MHz (blue) and 8907 MHz (red). b) NMR spectrum of 99\% $^{13}$C-enriched acetonitrile after accumulating 12676 scans. The diamond DNP signal corresponds to a polarization on the order of 6\%, an enhancement of $\sim$170,000 over thermal equilibrium. $^{13}$C nuclear polarization as a function of applied microwave frequency at the c) $m_s=0$ to $m_s=-1$ and d) $m_s=0$ to $m_s=+1$ NV$^-$ transitions. The opposite signs of these two curves are consistent with the opposite electron spin polarizations of the two NV$^-$ transitions. Data were acquired with a laser intensity of 16 $\frac{W}{cm^2}$ and microwave power of 1.3 W.}
 \end{figure}

 \begin{figure}[H]
 \includegraphics[width=0.5\textwidth]{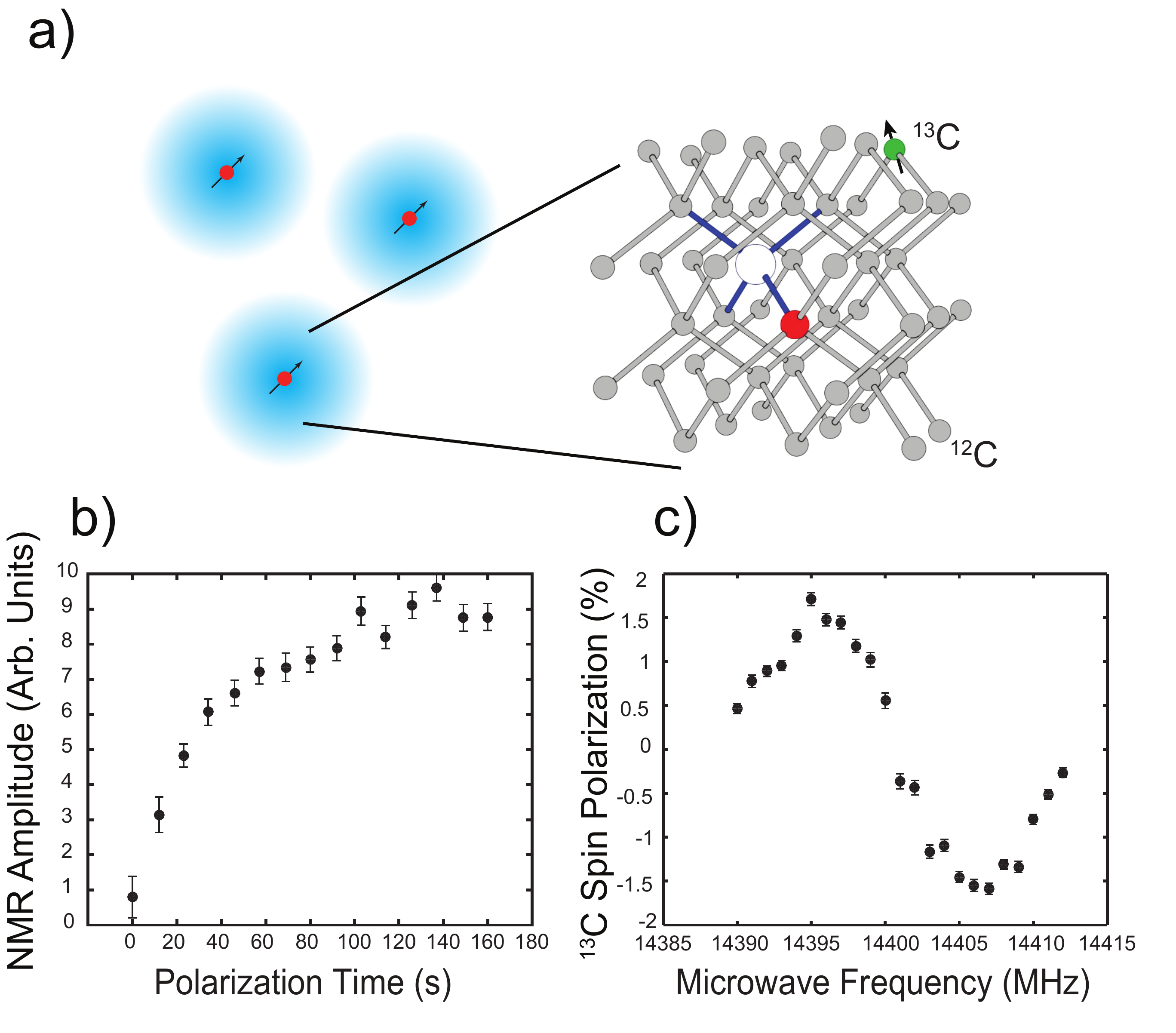}%
 \caption{\label{dnp}a) Schematic representation of the DNP process. Direct polarization near NV$^-$ centers (red) gives rise to $^{13}$C (green) spin hyperpolarization. Spin diffusion carries the polarization (blue) to the bulk material until a steady-state is reached. b) Time dependence of $^{13}$C spin polarization obtained by $^{13}$C NMR at 4.5 MHz. Beyond approximately 100 seconds the spin polarization has reached a steady state that represents a balance between the hyperpolarization/spin diffusion process and the spin lattice relaxation of the nuclear spins.}
 \end{figure}
 


These results introduce a methodology for hyperpolarization of diamond samples that we anticipate can be efficiently translated to crafted high surface area diamond crystals and nanocrystals in arbitrary magnetic fields and crystal orientations, thereby providing a source of hyperpolarization that can be transferred to solid and liquid samples under ambient and near-ambient conditions. We envision highly enhanced NMR of liquids and solids, including freeze thaw cycles under mild variable temperature conditions to enable solid-state polarization transfer for liquid samples while maintaining compatibility with biological systems. Hyperpolarization of samples in contact with diamonds and subject to solid-state cross polarization \cite{alex} or liquid-state cross relaxation \cite{slichter} represents a robust extension of polarization transfer mechanism previously demonstrated from hyperpolarized $^{129}$Xe to solid \cite{xesolid} and liquid samples \cite{spinoe}.


\section*{}
This work was supported by the Director, Office of Science, Office of Basic Energy
Sciences, Materials Sciences and Engineering Division, of the US Department of Energy
under Contract No. DE-AC02-05CH11231. K. J. acknowledges fellowship support from the Republic of Korea Army. The authors thank Eric Scott and Melanie Drake for providing the diamond sample used in this study. The authors also thank Jeffrey Reimer, Birgit Hausmann and Carlos Meriles for helpful discussions.

\bibliography{dnpbib}{}

\begin{thebibliography}{10}

\bibitem{raftery}
D.~Raftery, {\em {Xenon NMR spectroscopy}}, vol.~{57} of {\em {Annual Reports
  on NMR Spectroscopy}}.
\newblock {Elsevier Academic Press Inc.}, {2006}.

\bibitem{happer}
T.~Walker and W.~Happer {\em {Rev. Mod. Phys.}}, vol.~{69}, pp.~{629--642}, {}
  {1997}.

\bibitem{schroeder}
C.~Witte and L.~Schroeder {\em {NMR in Biomedicine}}, vol.~{26},
  pp.~{788--802}, {} {2013}.

\bibitem{carver}
M.~Bouchiat, T.~Carver, and C.~Varnum {\em {Phys. Rev. Lett.}}, vol.~{5},
  no.~{8}, pp.~{373--375}, {1960}.

\bibitem{jefftrends}
J.~A. Reimer {\em {Solid State NMR}}, vol.~{37}, pp.~{3--12}, {} {2010}.

\bibitem{hayesreview}
S.~E. Hayes, S.~Mui, and K.~Ramaswamy {\em {J. Chem. Phys.}}, vol.~{128}, {}
  {2008}.

\bibitem{mewis}
S.~B. Duckett and R.~E. Mewis in {\em {Hyperpolarization Methods in NMR
  Spectroscopy}} ({Kuhn, LT}, ed.), vol.~{338} of {\em {Topics in Current
  Chemistry}}, pp.~{75--103}, {}: {Springer-Verlag Berlin}, {2013}.

\bibitem{weitekamp}
C.~Bowers and D.~Weitekamp {\em {Phys. Rev. Lett.}}, vol.~{57},
  pp.~{2645--2648}, {} {1986}.

\bibitem{bargon}
J.~Natterer and J.~Bargon {\em {Progress in Nuclear Magnetic Resonance
  Spectroscopy}}, vol.~{31}, pp.~{293--315}, {} {1997}.

\bibitem{griffin}
Q.~Z. Ni, E.~Daviso, T.~V. Can, E.~Markhasin, S.~K. Jawla, T.~M. Swager, R.~J.
  Temkin, J.~Herzfeld, and R.~G. Griffin {\em {Accounts of Chemical Research}},
  vol.~{46}, pp.~{1933--1941}, {} {2013}.

\bibitem{prisner}
C.~Griesinger, M.~Bennati, H.~M. Vieth, C.~Luchinat, G.~Parigi, P.~Hoefer,
  F.~Engelke, S.~J. Glaser, V.~Denysenkov, and T.~F. Prisner {\em {Progress in
  Nuclear Magnetic Resonance Spectroscopy}}, vol.~{64}, pp.~{4--28}, {} {2012}.

\bibitem{larsen}
J.~Ardenkjaer-Larsen, B.~Fridlund, A.~Gram, G.~Hansson, L.~Hansson, M.~Lerche,
  R.~Servin, M.~Thaning, and K.~Golman {\em {PNAS}}, vol.~{100},
  pp.~{10158--10163}, {} {2003}.

\bibitem{bodenhausen}
D.~Gajan, A.~Bornet, B.~Vuichoud, J.~Milani, R.~Melzi, H.~A. van Kalkeren,
  L.~Veyre, C.~Thieuleux, M.~P. Conley, W.~R. Gruening, M.~Schwarzwaelder,
  A.~Lesage, C.~Coperet, G.~Bodenhausen, L.~Emsley, and S.~Jannin {\em {PNAS}},
  vol.~{111}, pp.~{14693--14697}, {} {2014}.

\bibitem{goez}
M.~Goez {\em {Concepts in Magnetic Resonance}}, vol.~{7}, pp.~{69--86}, {}
  {2005}.

\bibitem{triplet}
K.~Tateishi, M.~Negoro, S.~Nishida, A.~Kagawa, Y.~Morita, and M.~Kitagawa {\em
  {PNAS}}, vol.~{111}, pp.~{7527--7530}, {} {2014}.

\bibitem{rugar}
D.~Rugar, H.~J. Mamin, M.~H. Sherwood, M.~Kim, C.~T. Rettner, K.~Ohno, and
  D.~D. Awschalom {\em {Nat. Nano}}, 2014.

\bibitem{devince}
S.~J. DeVience, L.~M. Pham, I.~Lovchinsky, A.~O. Sushkov, N.~Bar-Gill,
  C.~Belthangady, F.~Casola, M.~Corbett, H.~Zhang, M.~Lukin, H.~Park,
  A.~Yacoby, and R.~L. Walsworth vol.~arXiv:1406.3365 [quant-ph], 2014.

\bibitem{degen}
M.~Loretz, T.~Rosskopf, J.~M. Boss, S.~Pezzagna, J.~Meijer, and C.~Degen {\em
  {Science}}, {} {2014, (10.1126/science.1259464)}.

\bibitem{jelezko}
C.~Mueller, X.~Kong, J.~M. Cai, K.~Melentijevic, A.~Stacey, M.~Markham,
  D.~Twitchen, J.~Isoya, S.~Pezzagna, J.~Meijer, J.~F. Du, M.~B. Plenio,
  B.~Naydenov, L.~P. McGuinness, and F.~Jelezko {\em {Nat. Comm.}}, vol.~{5},
  {} {2014}.

\bibitem{sushkov}
A.~Sushkov, I.~Lovchinsky, N.~Chisholm, R.~Walsworth, H.~Park, and M.~Lukin
  {\em {Phys. Rev. Lett.}}, vol.~{113}, p.~{197601 (5 pp.)}, {} {2014}.

\bibitem{fischerensemble}
R.~Fischer, A.~Jarmola, P.~Kehayias, and D.~Budker {\em {Phys. Rev. B.}},
  vol.~{87}, {} {2013}.

\bibitem{haijing}
H.-J. Wang, C.~S. Shin, C.~E. Avalos, S.~J. Seltzer, D.~Budker, A.~Pines, and
  V.~S. Bajaj {\em {Nat. Comm}}, vol.~{4}, {} {2013}.

\bibitem{frydmanbulk}
R.~Fischer, C.~O. Bretschneider, P.~London, D.~Budker, D.~Gershoni, and
  L.~Frydman {\em {Phys. Rev. Lett.}}, vol.~{111}, {} {2013}.

\bibitem{alvarez}
G.~Alvarez, C.~Bretschneider, R.~Fischer, P.~London, H.~Kanda, J.~G.~D. Onoda,
  S.~Isoya, and L.~Frydman vol.~arXiv:1412.8635 [quant-ph], 2014.

\bibitem{king}
J.~P. King, P.~J. Coles, and J.~A. Reimer {\em {Phys. Rev. B.}}, vol.~{81}, {}
  {2010}.

\bibitem{carlos}
D.~Abrams, M.~E. Trusheim, D.~R. Englund, M.~D. Shattuck, and C.~A. Meriles
  {\em {Nano Letters}}, vol.~{14}, pp.~{2471--2478}, {} {2014}.

\bibitem{reynhardt}
E.~Reynhardt and G.~High {\em {J. Chem. Phys.}}, vol.~{109}, pp.~{4090--4099},
  {} {1998}.

\bibitem{abragam}
A.~{Abragam}, {\em Principles of Nuclear Magnetism}.
\newblock Oxford: Oxford University Press, 1961.

\bibitem{alex}
A.~Pines, M.~Gibby, and J.~Waugh {\em {J. Chem. Phys}}, vol.~{59}, no.~{2},
  pp.~{569--590}, {1973}.

\bibitem{slichter}
C.~P. {Slichter}, {\em Principles of Magnetic Resonance}.
\newblock New York: Springer, 3rd~ed., 1996.

\bibitem{xesolid}
H.~Gaede, Y.~Song, R.~Taylor, E.~Munson, J.~Reimer, and A.~Pines {\em {Applied
  Magnetic Resonance}}, vol.~{8}, no.~{3-4}, pp.~{373--384}, {1995}.

\bibitem{spinoe}
G.~Navon, Y.~Song, T.~Room, S.~Appelt, R.~Taylor, and A.~Pines {\em {Science}},
  vol.~{271}, pp.~{1848--1851}, {} {1996}.

\bibitem{dimaspaper}
V.~M. Acosta, E.~Bauch, M.~P. Ledbetter, C.~Santori, K.~M.~C. Fu, P.~E.
  Barclay, R.~G. Beausoleil, H.~Linget, J.~F. Roch, F.~Treussart,
  S.~Chemerisov, W.~Gawlik, and D.~Budker {\em {Phys. Rev. B.}}, vol.~{80}, {}
  {2009}.

\bibitem{poulis}
J.~Vanhouten, W.~T. Wenckebach, and N.~J. Poulis {\em Physica B and C},
  vol.~92, no.~2, pp.~210--220, 1977.

\bibitem{vega}
Y.~Hovav, A.~Feintuch, and S.~Vega {\em {Phys. Chem. Chem. Phys.}}, vol.~{15},
  no.~{1}, pp.~{188--203}, {2013}.

\end{thebibliography}
\bibliographystyle{ieeetr}

\section*{Supplemental Material}
\renewcommand{\thefigure}{S\arabic{figure}}
\setcounter{figure}{0}

\section{Methods}
In order to investigate dynamic nuclear polarization (DNP) effects with nitrogen vacancy (NV$^-$) centers, we constructed a combined dynamic nuclear polarization/optically-detected magnetic resonance/nuclear magnetic resonance instrument, shown schematically in Fig. S1. The magnetic field is supplied by a custom-built electromagnet (Tel-Atomic) and is set to 420 mT. A Coherent Verdi G15 laser delivers 532 nm illumination to the sample through a Gaussian beam with a waist of 1.5 mm, essentially illuminating the entire surface of the diamond with an intensity up to 16 $\frac{W}{cm^2}$. Fluorescence is separated from excitation light by a dichroic mirror and  detected by an avalanche photodiode. Optically-detected magnetic resonance (ODMR) is performed by monitoring the diamond fluorescence while varying the applied microwave frequency. Microwave irradiation is delivered to the sample by a microwave loop of diameter 9.6 mm. NMR was performed using a Magritek Kea 2 spectrometer  with a homebuilt 50-turn planar coil probe tuned to $\sim4.5$ MHz.

A commercially-available $2\times2\times0.32$ mm, $\langle 100\rangle$ surface-orientation single crystal of synthetic high-pressure, high-temperature diamond (Sumitomo) was acquired. Electron irradiation at 1 MeV with a fluence of $10^{18}$cm$^{-2}$ followed by annealing at 800\degree C yielded an ensemble of NV$^-$ centers. NV$^-$ concentration under these conditions is expected to be on the order of $10^{18}$cm$^{-3}$\cite{dimaspaper}. The crystal was mounted on a goniometer inside the electromagnet, and one of the $\langle111\rangle$ axes was aligned with the magnetic field by monitoring the ODMR spectrum. In this orientation, there are three equivalent ODMR spectra of the NV$^-$ centers along $\langle 111\rangle$  axes at an angle of 109.5$\degree$ with respect to the magnetic field and a single ODMR spectrum corresponding to the aligned NV$^-$ centers. With the field set to $420$ mT, ODMR and DNP were performed using microwave fields at $8,900$ MHz and $ 14,600$ MHz. For the misaligned NV$^-$ data, the sample holder/NMR probe was rotate 90 degrees around the vertical axis, which is perpendicular to both the magnetic field and laser. In this configuration, the relative orientation of the NMR coil and magnetic field is identical to the ``aligned" measurements, and the same experimental parameters are valid. A separate reference measurement (described later) was performed for this configuration in case of unintended variations of the NMR sensitivity.

DNP data were acquired by polarizing for 60 s unless otherwise noted. Then, a $\frac{\pi}{2}$ NMR pulse of duration 10 $\mu$s generated transverse magnetization that was inductively detected. Time-domain NMR data were apodized by exponential multiplication with a decay constant of 1 ms. After application of phase correction and a fast Fourier transform algorithm, frequency-domain spectra were fitted to single Lorentzian functions. The nuclear polarization was taken to be proportional to the amplitude of the fitted peak; error bars represent the 95\% confidence intervals for the amplitude. The polarization buildup was monitored with a saturation recovery pulse sequence, where the polarization was initially destroyed with a series of $\frac{\pi}{2}$ pulses followed by a variable polarization time and NMR detection. The buildup data were processed separately by exponential apodization and Fourier transform with Magritek Prospa software, followed by phase correction and fitting to Lorentzian functions to extract the amplitude.

\section{Validity of External Calibration}
Typically, when performing quantitative NMR studies a reference sample of known quantity is mixed with the sample and observed under identical conditions. However, observing $^{13}$C NMR at 420 mT is inherently difficult due to the low sensitivity, and without hyperpolarization it is impossible to observe an NMR signal with the same number of scans as the DNP experiments. We therefore used a 99\% $^{13}$C enriched sample of dimethyl sulfoxide, doped with gadolinium (III) to achieve a spin-lattice relaxation time less than 2 ms. This allowed accumulation the of more than 10,000 scans needed to achieve a sufficient signal-to-noise ratio for the thermally-polarized liquid whose polarization was $3.5\times10^{-7}$.

In order for the calibration to be accurate, several conditions must be met. First, all the parameters of the NMR experiment must remain constant, including pulse parameters and the quality factor of the resonant probe. Additionally, the reference sample should match the diamond in shape and be positioned identically with respect to the NMR coil to avoid miscalibration from the inhomogeneous sensitivity of the NMR coil. Care was taken to maintain all experimental parameters, but the problem of inhomogeneous sensitivity requires additional consideration. As it was impractical to use a liquid sample with identical dimensions as the diamond (2 mm $\times$ 2 mm $\times$ 0.32 mm), a cylindrical liquid sample of diameter 3.7 mm and approximate depth 2 mm was used. Assuming the pulse parameters are calibrated to give a $\frac{\pi}{2}$ rotation at the center of the coil, the rotation angle was estimated as a function of position by numerically simulating the magnetic field (B$_1$) of the RF coil. Sensitivity to sample magnetization is also proportional to B$_1$, and these values were combined to create a map of sensitivity as function of position (Fig. S2). From these simulations it was estimated that the NMR sensitivity for the DMSO liquid reference sample was approximately 99$\%$ that of the diamond sample, and we conclude that RF field inhomogeneities do not cause significant error in the calibration.

There are a variety of other possible sources of error in the calibration of the diamond  $^{13}$C polarization, including the measurement of the number of spins in the diamond and DMSO sample, the deviation of the coil sensitivity from the numerical simulations, and possible drifts in the quality factor of the probe. However, the primary source of error is expected to be the noise relative to the weak calibration signal. This noise was characterized by fitting the DMSO spectrum with a Lorentzian function as described earlier and obtaining an error estimate of 6.7\%. This error applies uniformly to the calibration of all DNP data and is thus not represented as individual error bars. This possible error does not affect the conclusions of this study.


\section{Mechanisms of Dynamic Nuclear Polarization}
The mechanisms for DNP via fixed paramagnets in insulating solids are well known and described elsewhere \cite{slichter,reynhardt,abragam}. We briefly review the relevant mechanisms here. For a diamond containing $^{13}$C and NV$^-$ spins, there exist transitions at frequencies $\omega_{NV}\pm\omega_{^{13}C}$ involving simultaneous nuclear and electron spin flips that are nominally forbidden in the absence of electron-nuclear coupling. Dipolar coupling of these spins creates a non-zero transition probability, and a sufficiently strong microwave field may then drive the transition. These transitions are either zero quantum (induced by coupling terms of the form $S^\pm I^\mp$) or double quantum ($S^\pm I^\pm$) and may be selected by frequency. Since optical pumping of the NV$^-$ center populates the $m_s=0$ states, each of these transitions is polarized and the effect of driving the forbidden transitions is to preferentially induce nuclear spin flips of a particular sign. Thus, the net result of the combined optical pumping and microwave irradiation is to preferentially populate one of the nuclear spin states. This is known as the the ``solid effect" method of DNP, since it relies on the dipolar interaction of fixed spins in solids \cite{reynhardt,slichter,abragam}.
 
If, however, the NV$^-$ spin transitions are homogeneously broadened by their mutual dipole-dipole interactions, as is the case with high NV$^-$ concentrations, then there exist energy-conserving transitions that involve a nuclear spin flip accompanied by multiple NV$^-$ spin flips. In this case the ``dipolar energy reservoir" of the NV$^-$ centers is in thermal contact with the nuclear spins. This coupling can provide a relaxation pathway for the nuclear spins or, if the dipolar ``spin temperature" is perturbed by optical pumping or microwave saturation, it can provide a second method of DNP known as ``thermal mixing" \cite{poulis,king,reynhardt}. Application of microwave irradiation connects the dipolar energy reservoir to the electron Zeeman energy in the rotating frame \cite{vega}, and thus provides a pathway for electron spin polarization transfer to nuclei. The solid effect and thermal mixing DNP have similar dependencies on microwave frequency, where the maximum polarization occurs approximately $\pm\omega_{^{13}C}$ offset from the center of the ODMR transition. The results presented here likely contain contributions from both solid-state effect and thermal mixing effects. We expect the relative magnitudes of these effects to be sample dependent, owing to differing NV$^-$ concentration.


\section{Microwave and Laser Power Dependence}
The effectiveness of the OP/DNP process was found to increase slightly with laser intensity. While the exact relationship between laser intensity and NV$^-$ spin polarization under these conditions is unknown, it is clear that optical absorption is significant over the depth of the diamond. The optical absorption coefficient in the diamond is approximately 9 mm$^{-1}$, and at thickness of 0.32 mm, only 6$\%$ of the light is transmitted. Increasing the laser light then effectively increases the volume of the sample that is highly polarized as well as possibly increasing the degree of polarization near the surface of the sample. The  Fig. S3 shows DNP data as a function of laser intensity.

The effectiveness of OP/DNP also increases with applied microwave power (Fig. S4). This is expected for DNP in diamond, where the solid effect or thermal mixing mechanisms of DNP are expected to contribute. In the case of the solid-effect, the increased microwave power will more effectively drive the ``forbidden transitions" involving mutual spin flips of $^{13}$C and NV$^-$ spins. For thermal mixing, the microwaves drive the NV- dipole energy reservoir into equilibrium with rotating frame spin temperature of the NV$^-$ centers. In each case, stronger microwave irradiation results in more efficient transfer of polarization.

   \begin{figure}[h]
 \includegraphics[width=0.5\textwidth]{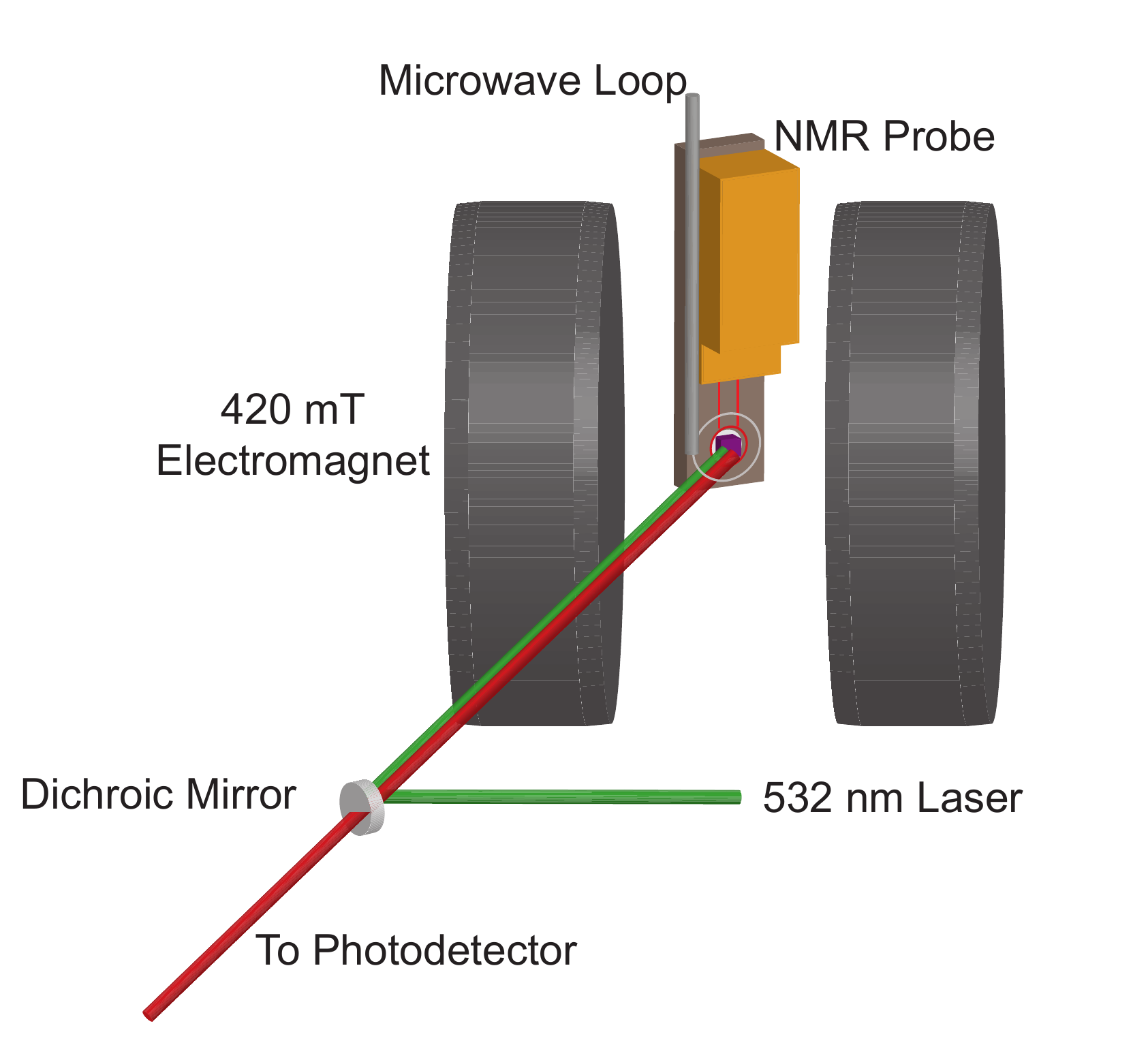}%
 \caption{\label{setup}Experimental setup for optically detected magnetic resonance, dynamic nuclear polarization, and nuclear magnetic resonance. An unfocused 532 nm laser beam provides optical pumping. A tuned NMR probe provides RF irradiation and detection while a microwave loop provides microwave irradiation.}
 \end{figure}

   \begin{figure}[h]
 \includegraphics[width=0.5\textwidth]{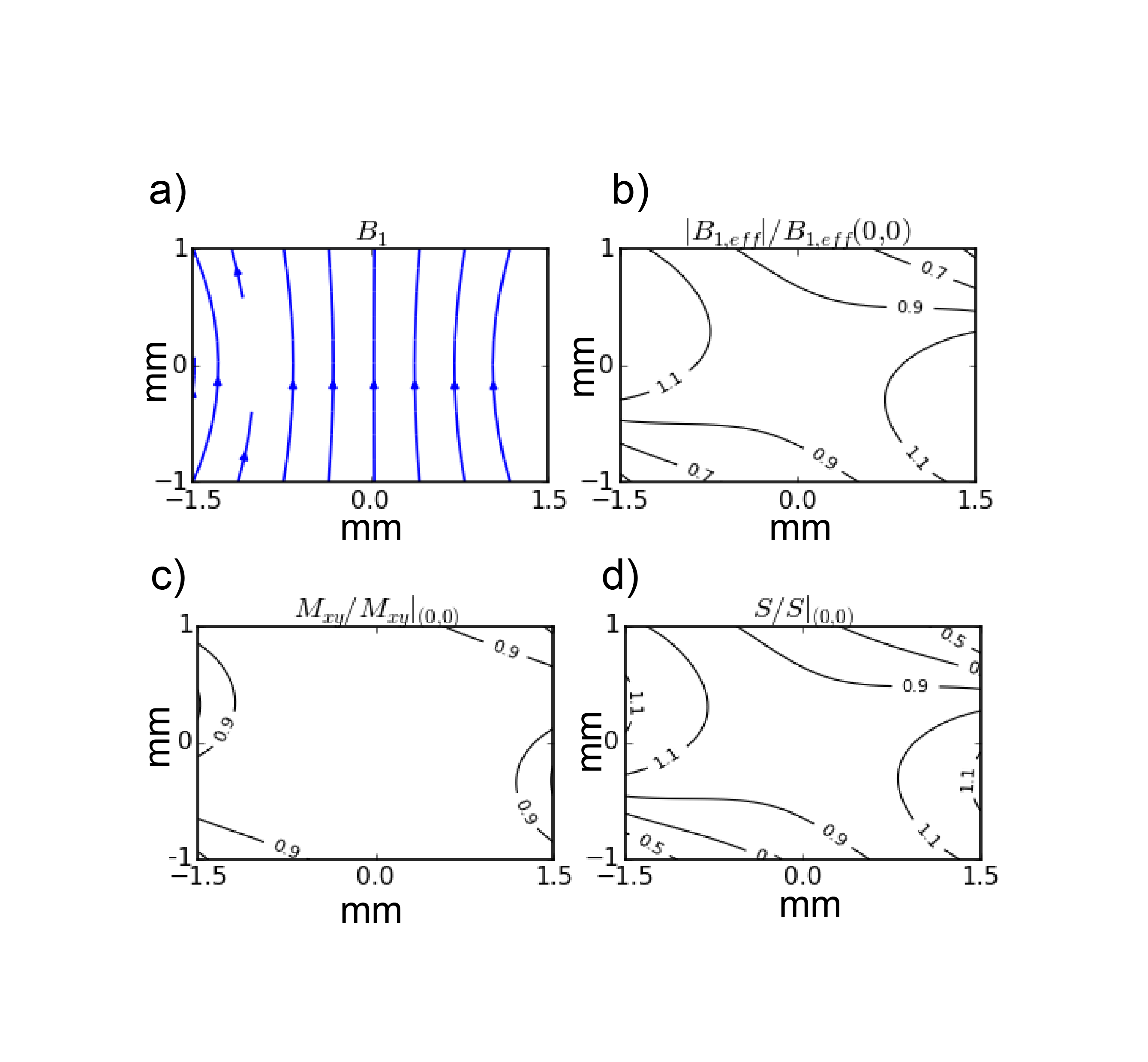}%
 \caption{\label{setup}a) Cross-section of field lines for RF pulses from planar NMR coil. b) Effective RF field perpendicular to the static magnetic field oriented 45$\degree$ to the coil axis. c) Normalized nuclear magnetization generated by a nominal $\frac{\pi}{2}$ pulse. d) Overall NMR sensitivity as a function of position, accounting for generation of transverse nuclear magnetization and inductive sensitivity of the coil.}
 \end{figure}

    \begin{figure}[h]
 \includegraphics[width=0.5\textwidth]{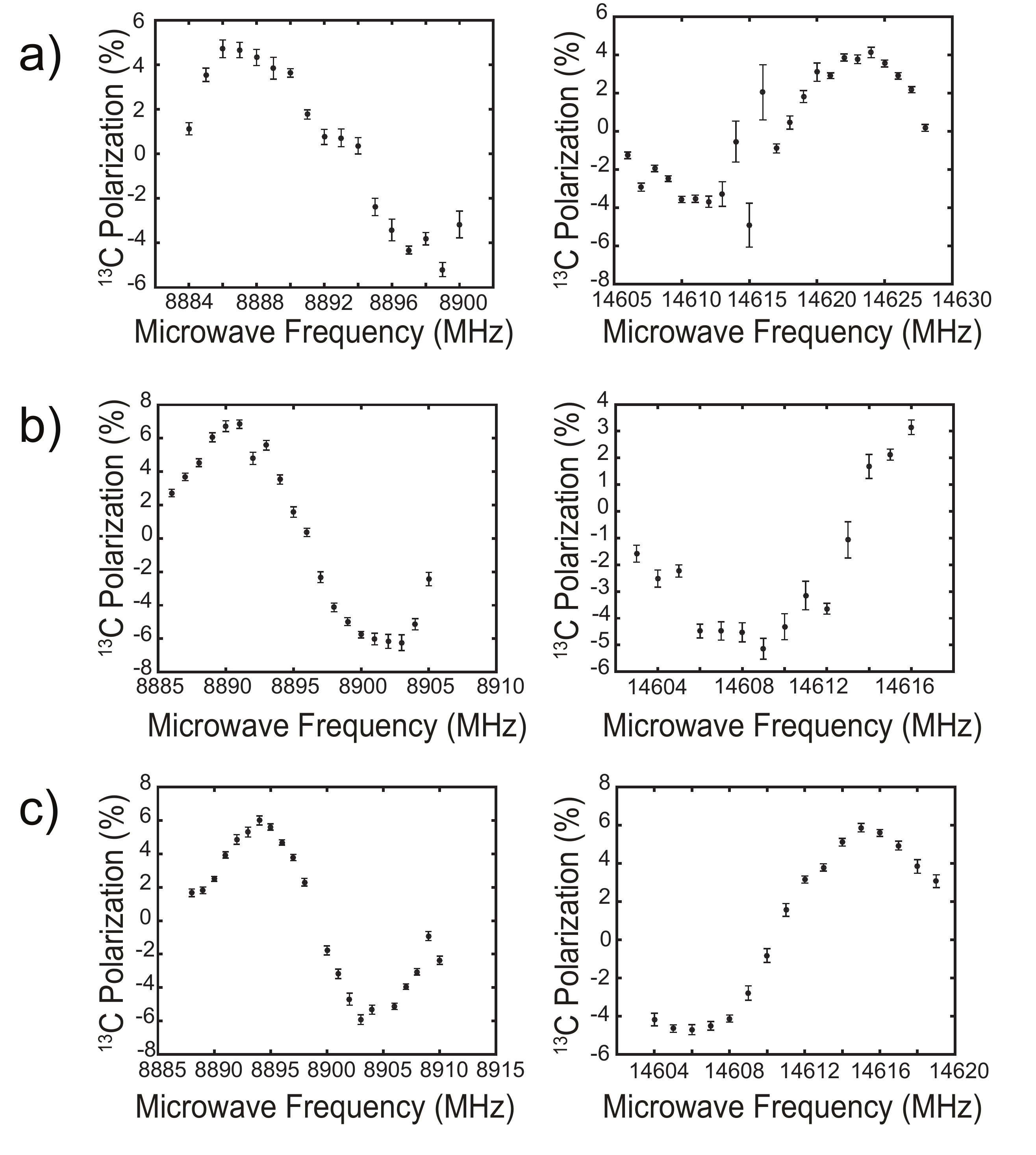}%
 \caption{\label{S3}DNP curves obtained with varying laser intensity: a) 11 $\frac{W}{cm^2}$, b) 16 $\frac{W}{cm^2}$, c) 45 $\frac{W}{cm^2}$. Increased laser intensity results in a slight increase in nuclear polarization.}
 \end{figure}

    \begin{figure}[h]
 \includegraphics[width=0.5\textwidth]{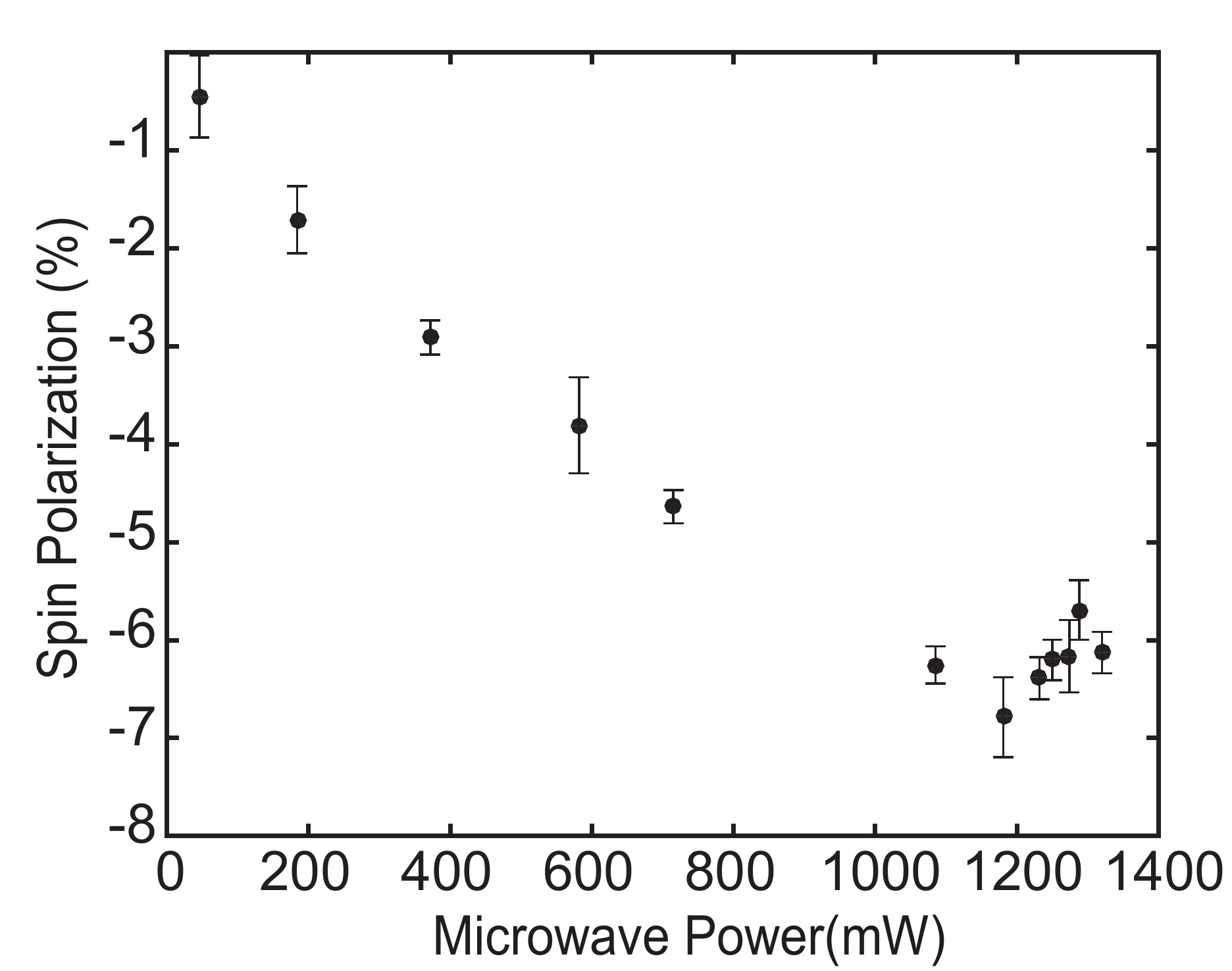}%
 \caption{\label{power}Nuclear spin polarization as a function of microwave power after 60 s of DNP at 8896 MHz and  16 $\frac{W}{cm^2}$ illumination.}
 \end{figure}



  


\end{document}